\documentclass[10pt,oneside,twocolumn,letterpaper,journal]{IEEEtran}
\usepackage{verbatim}
\usepackage{xcolor}
\usepackage{multirow}

\newcommand\hlbreakable[1]{\textcolor{black}{#1}}

\usepackage[T1]{fontenc}
\usepackage{booktabs}

\usepackage{cite}

\usepackage{graphicx}
\usepackage{adjustbox}
\ifCLASSINFOpdf
\else
\fi

\usepackage{amsmath}

\usepackage[cmintegrals]{newtxmath}
\usepackage{algorithm}
\usepackage{algorithmic}

\usepackage{subfig}
\usepackage[displaymath]{lineno}

\begin{document}
\title{Knowledge Distillation Based Semantic Communications For Multiple Users}

\author{Chenguang Liu, Yuxin Zhou, Yunfei Chen, {\em Senior Member, IEEE},\\ Shuang-Hua Yang, {\em Senior Member, IEEE}
      \thanks{Chenguang Liu is with the School of Engineering, University of Warwick, Coventry, UK, CV4 7AL. {e-mail: Chenguang.Liu@warwick.ac.uk}
      }
      \thanks{Yuxin Zhou is with the Department of Computer Science and Engineering, Southern University of Science and Technology, Shenzhen, China. {e-mail: zhouyx2020@mail.sustech.edu.cn}.}
      \thanks{Yunfei Chen is with the Department of Engineering, University of Durham, Durham, UK, DH1 3LE. {e-mail: Yunfei.Chen@durham.ac.uk}.}
      \thanks{Shuang-Hua Yang is with Shenzhen Key laboratory of Safety and Security for Next Generation of Industrial Internet, Southern University of Science and Technology, Shenzhen, China, and also with Department of Computer Science, University of Reading, UK. {e-mail: yangsh@sustech.edu.cn}.}
  }

\markboth{Journal of \LaTeX\ Class Files,~Vol.~14, No.~8, August~2015}%
{Liu \MakeLowercase{\textit{et al.}}: Knowledge based semantic communications for multiple users}

\maketitle
\begin{abstract}
Deep learning (DL) has shown great potential in revolutionizing the traditional communications system. Many applications in communications have adopted DL techniques due to their powerful representation ability. However, the learning-based methods can be dependent on the training dataset and perform worse on unseen interference due to limited model generalizability and complexity. In this paper, we consider the semantic communication (SemCom) system with multiple users, where there is a limited number of training samples and unexpected interference. To improve the model generalization ability and reduce the model size, we propose a knowledge distillation (KD) based system where Transformer based encoder-decoder is implemented as the semantic encoder-decoder and fully connected neural networks are implemented as the channel encoder-decoder. Specifically, four types of knowledge transfer and model compression are analyzed. Important system and model parameters are considered, including the level of noise and interference, the number of interfering users and the size of the encoder and decoder. Numerical results demonstrate that KD significantly improves the robustness and the generalization ability when applied to unexpected interference, and it reduces the performance loss when compressing the model size.  

\end{abstract}


\IEEEpeerreviewmaketitle

\section{Introduction}
\IEEEPARstart{A}{ccording} to Shannon and Weaver \cite{shannon}, communications can be categorized into three levels: transmission of symbols, transmission of semantics behind symbols, and effectiveness of semantics transmission. The first level aims to accurately transmit the symbols from the transmitter to the receiver by minimizing the bit error rate (BER) or the symbol error rate (SER). The second level semantic communication (SemCom) focuses on precisely conveying the meaning behind the bits. The third level concentrates on the effectiveness of the tasks that the communication intends to achieve over semantics transmission.




\hlbreakable{However, the limited spectrum resources constrain the capacity of traditional data communications at the first level, following the Shannon limit. To address this, SemCom extracts the meaning behind data and transmits only the essential semantic information, prioritizing semantic-level fidelity over bit-level accuracy. This is useful for applications requiring extensive data exchange but with limited bandwidth, where task effectiveness precedes exact information recovery. Potential applications include human-machine symbiosis, intelligent transportation, and extended reality (XR) \cite{9955312, 9663101, qin2022semantic}. For instance, the XR performance relies on processing the essential user data (\emph{e.g.}, head movement, gestures and text input). Fast data transmission via low-latency networks to XR servers is vital for data processing and the corresponding tactile feedback. By filtering out the non-essential data with semantic understanding, SemCom allows end devices to transmit only pertinent data required for the operation at the XR server, thereby reducing bandwidth requirement and computational costs on the XR server. To enable such functions, it is crucial to investigate effective techniques for extracting semantic information.}

Recently, deep learning (DL) has been widely applied to address problems in natural language processing and computer vision due to their powerful pattern recognition and representation capacity. Inspired by this, several works have been conducted to explore the DL-enabled SemCom systems for text\cite{8461983,9322296,9398576,9252948,9450827}, image\cite{8723589,9066966,9685036,9685667} and speech transmission\cite{9450827}. Different channel conditions are considered, including additive white Gaussian noise, Rician fading, and Rayleigh fading. However, it has yet to be studied whether a SemCom system, well-suited for end-to-end (E2E) communications\cite{9398576}, can effectively function in the presence of multi-user (MU) interference. MU interference, such as co-channel interference, is usually caused by multiple radios transmitting on the same frequency simultaneously due to the overly crowded spectrum\cite{939839}. Although interference can be avoided or mitigated through effective resource management\cite{6655494, 8070467}, these come at the expense of system complexity and resource utilization efficiency. When there is a high user density or uncontrolled interference sources, eliminating MU interference may not be practical. Allowing interference to co-exist with SemCom systems without significantly degrading the system performance could be a simple but effective way to overcome this challenge. Therefore, it is necessary to evaluate the SemCom system quantitatively and whether the learning-based techniques can achieve a tolerable performance in the presence of MU interference.

Moreover, the work in \cite{8839651} has emphasized that researchers should focus on not only applying the existing DL techniques to the improvement of the current communications system but also considering the requirements and constraints of the communications network, such as low model complexity and low power consumption for low-power chips, to enable learning and data-driven distributed mobile devices. The semantic information varies with the transmission task, such as a highly convoluted image feature map or compressed text embeddings\cite{9450827}. Therefore, we need a powerful model to mine the deeper information hidden in the raw training data and understand the relationship of the transmitted words. This typically requires a model designed with a complex and deep structure and an extensive amount of training data to cultivate the generalization ability. However, when it comes to deployment, a lighter model is preferred due to the constraints of computation complexity and time. To achieve this, knowledge distillation (KD) can be used to keep the light size and the generalizability, which was first proposed in \cite{44873} to compress the knowledge from an intelligent ensemble of models into a single light model. KD is widely utilized to reduce the size and improve the generalization for language understanding tasks\cite{Sanh2019DistilBERTAD,jiao-etal-2020-tinybert}. Nevertheless, KD has not yet been applied to SemCom.

\subsection{Related work}
To explore how learning-based algorithms can transform the communications system for lower complexity and better performance, representative works were conducted in \cite{8839651, 8054694,8663966, CALVANESESTRINATI2021107930}. The challenges and opportunities of machine learning in communications were reviewed and discussed in \cite{8839651} for the physical layer. The future research directions powered by the data-driven and learning approaches were pointed out. The recent advances in applying DL in the physical layer were demonstrated in \cite{8054694,8663966} to provide potential research directions for intelligent learning-based communications. The work in \cite{8227772} proposed a deep neural network for multi-input multi-output (MIMO) detection in different channel conditions, which has near-optimal performance with perfect channel state information (CSI). To address the channel distortion, a fully connected neural network was proposed for channel estimation and signal detection in the OFDM system, which has comparable system performance with the minimum mean-square error estimator\cite{8052521}. The work in \cite{8454325} demonstrated that the learning-based detector could perform without knowledge of CSI by proposing a sliding bidirectional recurrent neural network to detect the signals. Moreover, the effect of co-channel interference and radar interference on learning-based detectors were analyzed in \cite{LIU2021101343} and \cite{9733260}, respectively. 

Unlike the aforementioned works that only optimize and deploy the learning-based receiver, several works have been conducted to jointly optimize the transmitter and receiver. For example, an E2E learning-based communications system was proposed in \cite{8054694} using autoencoders to replace the traditional transmitter and receiver, to significantly reduce the complexity of design and implementation compared with traditional block-wise communications systems. Recently, research has shifted from transmission of symbols to transmission of semantic meaning inspired by the significant advancements of DL in natural language processing. The scalability and capacity of DL enables semantic understanding on deeper information, such as word meaning and word relations in text transmission, to improve the system performance. A comprehensive overview was conducted in \cite{CALVANESESTRINATI2021107930} on how the communications system can benefit from semantic and goal-oriented communications in terms of effectiveness and sustainability. This overview strengthens the idea that recovering the meaning behind the bits or completing the task that the transmission intends to achieve is key to the recovery of the transmitted information at the receiver. Understanding the meaning or the goals behind the bits requires that the coding and decoding schemes can identify the internal relationship of the transmitted information. 

There have been several studies on DL-enabled SemCom including text transmission \cite{8461983,9322296,9398576,9252948}, image transmission \cite{8723589,9066966,9685036,9685667}, speech transmission \cite{9450827} and task-oriented transmission \cite{9653664}. The work in \cite{8461983} proposed a joint source-channel coding (JSCC) communication framework using recurrent neural network for text transmission, which had lower word error rates than conventional coding schemes. By mapping the words in a semantic space, words with similar meaning can have close distance. Then, a DL-enabled SemCom system, DeepSC, was proposed in \cite{9322296,9398576} to use Transformer as the semantic encoder and decoder, to outperform traditional coding schemes, especially in low SNR regime. Additionally, a lite SemCom system, L-DeepSC, was proposed in \cite{9252948} for distributed IoT devices, which used parameter pruning and quantization to reduce the model size so that it can work with the limited bandwidth and transmission conditions. Moreover, the work in \cite{9450827} proposed an attention-based residual network as the joint transceiver for speech signals, which showed better robustness and performance than the traditional benchmarks with regard to the speech signal metrics. A JSCC was proposed for wireless image compression and transmission using two convolutional neural networks \cite{8723589}. The work in \cite{9066966} proposed a practical JSCC scheme based on autoencoder taking channel output feedback into account to improve the image reconstruction quality. The work in \cite{9685036} proposed an iterative source-channel decoder to explicitly consider residual bit error of each iteration for image transmission. The work in \cite{9685667} proposed coarse-to-fine image semantic coding model for multimedia SemCom system using generative adversarial networks. Apart from joint source-channel coding schemes for texts, images and speech, a task-oriented MU SemCom, MU-DeepSC, was designed to deal with multi-modal data \cite{9653664}. 

\subsection{Motivation and contribution}
Although all the previous works have demonstrated novelty and satisfactory performance by adopting DL-based semantic systems for robustness and effectiveness, there has not been any works on applying KD to the SemCom system with MU interference. MU interference of the communications system is inevitable in practice due to spectrum sharing. Yet, it is either not studied or ignored in the previous works. Moreover, revolutionizing the traditional communications system with a DL-enabled source-channel coding scheme still has to overcome many practical problems, including model generalizability and complexity. Specifically, the learning-based model can be overly dependent on the training dataset samples and consequently perform worse on unseen data. Several questions need to be addressed in DL-based SemCom systems:
\begin{enumerate}
\item How well can the model generalize on unseen interference?
\item How well can the model perform by training with limited dataset?
\item How light can the model be with negligible performance loss? 
\end{enumerate}
In this paper, our work focuses on the SemCom system using KD to improve the model generalization capacity and lower the model complexity. The main contributions of this paper can be summarized as below:
\begin{enumerate}
  \item To the best of the authors' knowledge, this is the first work that applies KD to the DL-enabled SemCom system with MU interference. The random occurrences and delays are considered for the interference. The performance of this system is evaluated for different signal-to-noise ratios (SNRs), signal-to-interference ratios (SIRs), and numbers of interfering users. 
  \item We propose four types of KD approaches by training the distilled student models for a limited range of SNR regimes in the absence of interference samples and then applying them to a wider range of SNR regimes with unseen MU interference. Numerical results show that distilled models outperform the non-distilled baselines and the conventional communications system with and without interference. Furthermore, it is proved that KD largely improves the generalizability and robustness of the model, which address Question 1) and 2) mentioned before. 
  \item By adopting model compression in KD after training, we apply model compression to the semantic encoder-decoder and the channel encoder-decoder to reduce performance loss. We also demonstrate the complexity and performance analysis in terms of the size and number of parameters, which address Question 3). Additionally, an ablation study is conducted to analyze the effect of various losses on the distilled student models.
\end{enumerate} 
The rest of the paper is organized as follows. In Section \ref{sec2}, we will introduce the system model of the SemCom system with MU interference and describe the main challenges. Section \ref{sec3} will discuss the KD-based SemCom, model compression and training process. Simulation settings and numerical results will be demonstrated in Section \ref{sec4}. Finally, Section \ref{sec5} will conclude the work.

Notation: To represent the parameters and outputs from the interfering user, we use the superscript $\mathcal{I}$ to represent the interfering user. Also, we use the superscript $\mathcal{T}$ and $\mathcal{S}$ to represent the parameters and outputs from the teacher model and the student model in the distillation training process. $\mathbb{E}$ denotes the mean of each element in the vector.

\section{System model and problem formulation} \label{sec2}
In this section, we will firstly describe the SemCom system with MU interference. Then, we will point out the challenges that DL-based SemCom might encounter, including generalizing on unseen data, limited training data and model complexity. 
\subsection{SemCom system}
\begin{figure}[!h]
  \centering
  \includegraphics[width=3.4in]{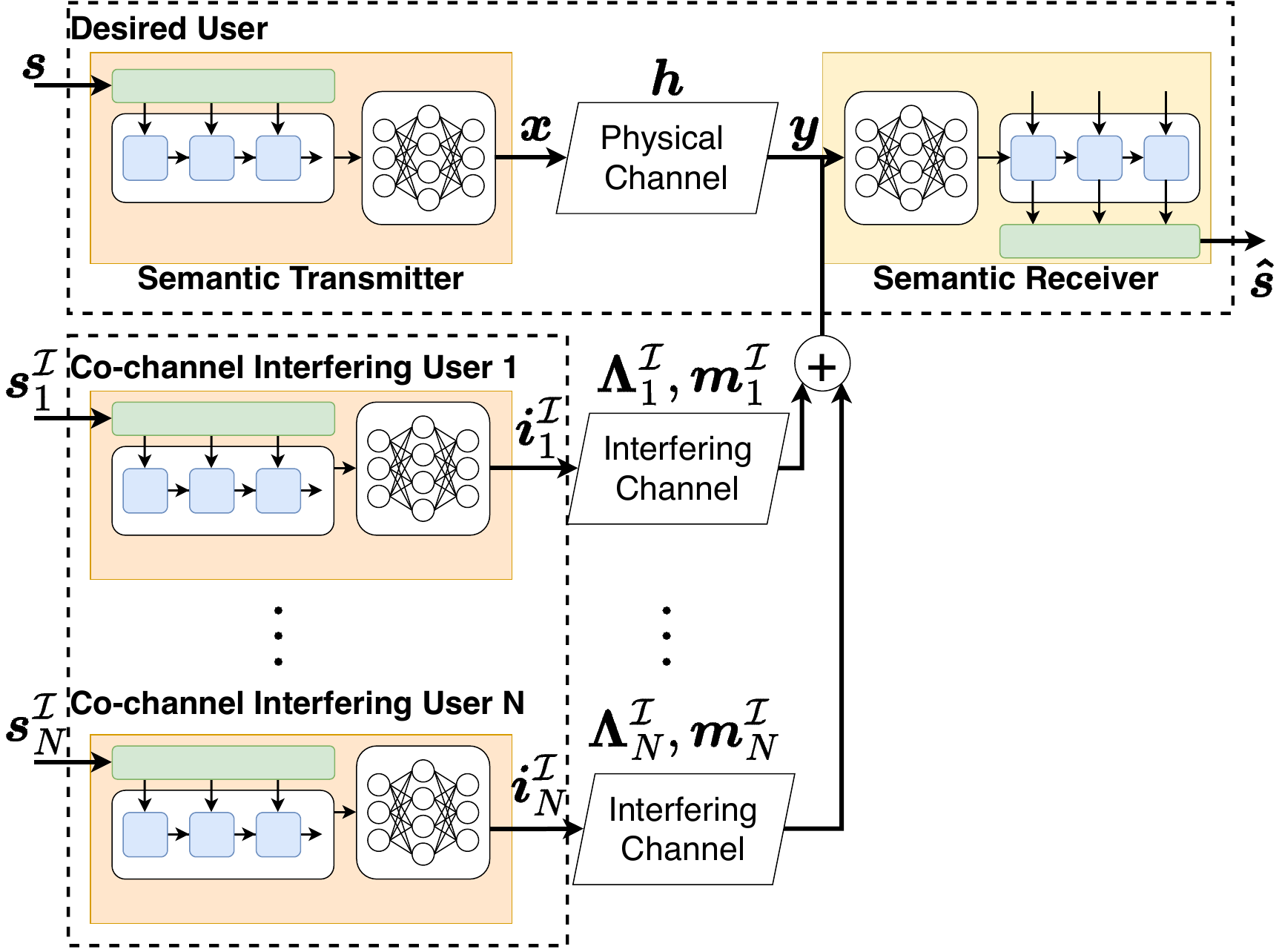}
  \caption{SemCom with MU interference due to co-channel competition.}
  \label{model}
\end{figure}
As shown in Fig. \ref{model}, we consider a SemCom system with multiple users, where all users are equipped with a semantic transmitter and compete for the same channel to incur co-channel interference. In this MU system, each user could be the desired user or cause interference to other users. When they transmit signals simultaneously, they can interfere with each other. Therefore, we model this SemCom system to have one desired user in the presence of co-channel interference from $N$ interfering users. Note that the interference in this paper refers to the signals transmitted from interfering users to the desired user. Each interference has random occurrences and delays in the transmission. Moreover, each desired user has a transmitter-receiver pair with one semantic encoder and channel encoder at the transmitter, one channel decoder and a semantic decoder at the receiver. The semantic encoder and decoder are responsible for compressing and extracting the information from the source at the semantic level. The channel encoder and decoder are designed to counteract the channel effect and recover the encoded semantic information.

We focus on a text transmission task using the SemCom system in the presence of MU interference and noise. The text input is expressed as $\boldsymbol{s}=[w_1, w_2,...,w_n]$, where $w_i$ denotes the $i$-th word in the sentence $\boldsymbol{s}$. Then, each word is successively encoded by the semantic encoder and channel encoder to formulate the transmitted symbols $\boldsymbol{x}$ as
\begin{align}
  \boldsymbol{p} & = \mathcal{SE}(\boldsymbol{s} ,\boldsymbol{\alpha}), \label{eq1} \\
  \boldsymbol{x} &= \mathcal{CE}( \boldsymbol{p},\boldsymbol{\beta}), \label{eq2}
\end{align}
where $\boldsymbol{s}$ is the text sentence as input, $\boldsymbol{p}$ denotes the semantic encoded information, $\mathcal{SE}(\cdot)$ and $\mathcal{CE}(\cdot)$ is the semantic encoder and the channel encoder with parameters $\boldsymbol{\alpha}$ and $\boldsymbol{\beta}$, respectively, and $\boldsymbol{x}$ is the transmitted symbols as output.

Consider a SemCom system with one desired user and multiple interfering users equipped with semantic transmitters, the interference signals transmitted by the $k$-th interfering user can be expressed by
\begin{equation}
  \boldsymbol{i}^\mathcal{I}_k =  \mathcal{CE}_{k}^\mathcal{I}(\mathcal{SE}_k^{\mathcal{I}}(\boldsymbol{s}_{k}^{\mathcal{I}},\boldsymbol{\alpha}_{k}^\mathcal{I}),\boldsymbol{\beta}_k^\mathcal{I}), \label{eq3}
\end{equation}
where $\boldsymbol{i}^\mathcal{I}_k$ is the interfering symbols from the $k$-th interfering user. Then, the signals transmitted by the interfering users arrive at the receiver of the desired user as,
\begin{gather}
  \boldsymbol{y} = \boldsymbol{h}*\boldsymbol{x} + \sum_{k=1}^N \mathbf{\Lambda}^\mathcal{I}_k * \boldsymbol{m}^\mathcal{I}_k * \boldsymbol{i}^\mathcal{I}_k + \boldsymbol{n}, \label{eq4}\\
  \mathbf{\Lambda}^\mathcal{I}_k = [\lambda_{k,1},\lambda_{k,2},\cdots, \lambda_{k,n}], \label{eq5}\\
  \lambda_{k,j} = \begin{cases}
       1 & \text{if the interference symbol occurs} \\
       0 & \text{otherwise}
    \end{cases} \label{eq6},
\end{gather}
where $\boldsymbol{h}$ and $\boldsymbol{m}^\mathcal{I}_k$ are the physical channel and the $k$-th interfering channel following Gaussian distributions, respectively, $\mathbf{\Lambda}_k$ denotes the random occurrence of the $k$-th interference, $\lambda_{k,n}$ is the binary interference occurrence indicator for $j$-th symbol in the $k$-th interference, $\boldsymbol{n}$ denotes the additive white Gaussian noise with mean zero and variance $\sigma^2$ and the operation $*$ denotes the element-wise multiplication. To decode the received signals for the desired user, the decoding process can be expressed by,
\begin{align}
  \boldsymbol{\hat{p}} & = \mathcal{CD}(\boldsymbol{y} ,\boldsymbol{\gamma}),\label{eq7} \\ 
  \boldsymbol{\hat{t}} &= \mathcal{SD}( \boldsymbol{\hat{p}},\boldsymbol{\delta}), \label{eq8}
\end{align}
where $\mathcal{CD}(\cdot)$ and $\mathcal{SD}(\cdot)$ are the channel decoder and semantic decoder with parameters $\boldsymbol{\gamma}$ and $\boldsymbol{\delta}$, respectively; $\boldsymbol{p}$ is the decoded channel information, which is also the input to the semantic decoder, $\boldsymbol{\hat{t}}$ denotes the decoded semantic information. Finally, a dense layer with softmax activation function is applied as the prediction layer to estimate the predicted sentence $\boldsymbol{\hat{s}}$ from the semantic decoded information $\boldsymbol{\hat{t}}$, which can be expressed by 
\begin{equation}
  \begin{split}
    \boldsymbol{\hat{s}} &= \mathcal{F}_{pred}(\boldsymbol{\hat{t}},\boldsymbol{w}_{pred},\boldsymbol{b}_{pred}) \\
  &= Softmax( \boldsymbol{w}_{pred}\boldsymbol{\hat{t}} + \boldsymbol{b}_{pred})
  \end{split} \label{eq9},
\end{equation}
where $\boldsymbol{w}_{pred}$ and $\boldsymbol{b}_{pred}$ are the prediction layer parameters, $\boldsymbol{\hat{s}}$ denotes the predicted sentence. 

The goal of the SemCom system is to recover the text sentence $\boldsymbol{\hat{s}}$ in the presence of interference and noise. In order to explore the generalization ability of the SemCom system, we assume the receiver has the perfect CSI of the desired-user channel gain $\boldsymbol{h}$ but no knowledge of the interfering channels. Then, the perfect CSI $\boldsymbol{h}$ is adopted by zero-forcing detector at the receiver to obtain the recovered signals $\boldsymbol{\hat{x}}$ from the received signal $\boldsymbol{y}$. The reason for this assumption is to focus on evaluating the proposed methods' performance in the presence of unseen interference rather than the effect of channel estimation errors or other practical limitations. The extension to the case with both interference and channel estimation error is not studied here due to space limit. Besides, we consider Rayleigh fading, random interference occurrences and transmission delay for the MU interference. 

\subsection{Problem description} \label{problem_description}
The cross-entropy loss is utilized to measure the difference between the ground truth hard labels and the predicted text sentence, which can be expressed by
\begin{equation}
  \begin{split}
  \mathcal{L}_{hard} =& \mathcal{L}_{CE}(\boldsymbol{s},\boldsymbol{\hat{s}}) \\
          =& -\frac{1}{n} \sum_{m=1}^n P(\boldsymbol{s}[m])\log P(\boldsymbol{\hat{s}}[m])
  \end{split}\label{eq10},
\end{equation}
where $\boldsymbol{s}=[w_1, w_2,...,w_n]$, $P(\boldsymbol{s}[m])$ is the probability for the real $m$-th word $w_m$ in the text sentence $\boldsymbol{s}$, and $P(\boldsymbol{\hat{s}}[m])$ is the probability for the predicted $m$-th word $\hat{w}_m$ in the text sentence $\boldsymbol{\hat{s}}$. Semantic transmitter and receiver are jointly optimized by adjust their parameters sets to minimize the loss considering physical channel attenuation, interference and noise. However, this training method also brings several challenges. 

The first challenge is the training of the semantic transmitter and receiver for generalization ability. \hlbreakable{The cross-entropy loss function only takes the final output of the SemCom system into account, so that the semantic encoder, channel encoder, channel decoder and semantic decoder are jointly trained as one black box. Although the SemCom system technically has a semantic transmitter and a semantic receiver, it is difficult to interpret the intermediate output as it is part of the convergence.} Consequently, the model can easily overfit on the training dataset or under-trained. This is inevitable for data-driven and learning-based systems, where the quality of the trained model cannot be guaranteed unless we conduct extensive experiments on all unseen data to validate its generalization ability. In practice, this may lead to excessive resources for training and testing. Therefore, it is important to design a training method so that the model can have considerable performance and generalization ability with unseen data. 

The second challenge is the limitation of the training data. The data-driven and learning-based communications system highly relies on the training dataset to maintain the model performance. An extensive amount of datasets containing sufficient patterns lays the foundation for training a powerful learning-based communications system. However, it can be challenging in a practical communications system to obtain the ideal dataset which meets such requirements. Therefore, it is important to train the model properly with fewer dataset.  

The third challenge is the model complexity. A complex model normally possesses a high convergence ability to accurately learn and approximate the relationship between inputs and outputs for a given dataset and generalizes well to unseen data. Nevertheless, the computational complexity can be too high for devices with limited computation resources. Therefore, it can be very challenging to reduce the model complexity while preserving its convergence. Next, we will address these challenges.

\section{KD-based SemCom system}\label{sec3}
In this section, we will introduce the Transformer-based semantic transceiver. Then, we will introduce a KD-based SemCom system to address the challenges mentioned above. Different model compression algorithms are used to address Question 3), and KD is adopted to solve Questions 1) and 2). Finally, the training procedure will be demonstrated.

\subsection{Transformer based semantic transceiver}
Inspired by the bidirectional encoder representations from Transformers (BERT) \cite{DBLP:conf/naacl/DevlinCLT19}, we adopt the Transformer structure as the semantic encoder and decoder to compress and extract the semantic information. The attention scheme of the Transformer can correlate the contextual information for each word in the sentence. The attention can be computed by
\begin{gather}
    \boldsymbol{Q} = \boldsymbol{W}^Q\boldsymbol{X}^Q,\quad \boldsymbol{K} = \boldsymbol{W}^K\boldsymbol{X}^K,\quad \boldsymbol{V} = \boldsymbol{W}^V\boldsymbol{X}^V, \label{eq11}\\
    \mathcal{F}_{A}(\boldsymbol{Q},\boldsymbol{K},\boldsymbol{V}) = Softmax(\frac{\boldsymbol{Q}\boldsymbol{K}^T}{\sqrt{d_k}})\boldsymbol{V}, \label{eq12}
\end{gather}
where $\boldsymbol{Q}$, $\boldsymbol{K}$, $\boldsymbol{V}\in\mathbb{R}^{B\times L\times D_{model}}$ denote the representations of the query, the key and the value with the input $\boldsymbol{X}^Q$, $\boldsymbol{X}^K$ and $\boldsymbol{X}^V\in\mathbb{R}^{B\times L\times D_{model}}$ and the parameter $\boldsymbol{W}^Q$, $\boldsymbol{W}^K$ and $\boldsymbol{W}^V$, respectively, $B$ is the batch size, $L$ is the sequence length and $D_{model}$ is the embedding dimension, $\mathcal{F}_{A}(\cdot)$ denotes the attention function and $d_k$ is the scaling factor. To obtain the information from different representation subspaces at different positions, the multi-head attention is used to calculate the attention in parallel and then concatenate the independent attention. The multi-head attention with $N$ heads can be computed by
\begin{gather}
  \boldsymbol{Q}_i = \boldsymbol{Q}\boldsymbol{W}^Q_i,\quad \boldsymbol{K}_i = \boldsymbol{K}\boldsymbol{W}^K_i, \quad \boldsymbol{V}_i = \boldsymbol{V}\boldsymbol{W}^V_i, \label{eq13}\\
  \mathcal{F}_{MA}(\boldsymbol{Q},\boldsymbol{K},\boldsymbol{V}) = [ \mathcal{F}_\mathcal{A}(\boldsymbol{Q}_1,\boldsymbol{K}_1,\boldsymbol{V}_1) ||\cdot\cdot\cdot|| \mathcal{F}_\mathcal{A}(\boldsymbol{Q}_{N},\boldsymbol{K}_{N},\boldsymbol{V}_{N}) ]\boldsymbol{W}^0, \label{eq14}
\end{gather}
where $\mathcal{F}_{MA}(\cdot)$ is the multi-head attention function with parameter $\boldsymbol{W}^0$, $\boldsymbol{Q}_i$, $\boldsymbol{K}_i$ and $\boldsymbol{V}_i\in \mathbb{R}^{B\times L\times D_{head}}$ are the query, the key and the value of the $i$-th head with parameters $\boldsymbol{W}^Q_i$, $\boldsymbol{W}^K_i$ and $\boldsymbol{W}^V_i \in \mathbb{R}^{D_{model}\times D_{head}}$, $D_{head} = D_{model} / N$ is the embedding dimension of each head and $||$ denotes the concatenation operation. Then, the feed forward layer is applied to the output of the multi-head attention layer, which is expressed as,
\begin{equation}
  \mathcal{F}_{FF}(\boldsymbol{X})= \boldsymbol{W}_\mathcal{FF}\boldsymbol{X} + \boldsymbol{b}_\mathcal{FF}, \label{eq15}
\end{equation}
where $\boldsymbol{W}_\mathcal{FF}$ and $\boldsymbol{b}_\mathcal{FF}$ are the parameters of the feed forward layer. Layer normalization is applied to each output from the multi-head attention layer and the feed forward layer to rescale and shift the outputs, which can be described by
\begin{equation}
    \mathcal{F}_{LN}(\boldsymbol{X}) = \frac{\boldsymbol{X}-\mathbb{E}[\boldsymbol{X}]}{\sqrt{\sigma_{\boldsymbol{X}}^2+\epsilon}}\theta+ \mu, \label{eq16} 
\end{equation}
where $\boldsymbol{X}\in \mathbb{R}^{D_{model}}$ is the input of the layer normalization, $\theta$ and $\mu$
are the trainable parameters, $\sigma_{\boldsymbol{X}}$ is the variance of the input $\boldsymbol{X}$, $\epsilon$ is 
an arbitrarily small number. Also, we apply skip connection using addition by adding the output from the preceding layer to the layer ahead.

For the Transformer based semantic transmitter, the text sequence $\boldsymbol{s}$ is preprocessed by a text tokenizer to split the text into words by punctuation and whitespaces, and then map each word with the corresponding scalar number according to the word representation dictionary. Then, the text sequence $\boldsymbol{s}$ is embedded as $\boldsymbol{t}\in\mathbb{R}^{B\times L\times D_{model}}$, which is the input to the Transformer layer. Each layer of the Transformer based semantic encoder contains a multi-head self-attention layer and a feedforward layer processed by residual connection and layer normalization. The Transformer layer of the semantic encoder can be expressed by,
\begin{align}
  \boldsymbol{z}_{self} &= \mathcal{F}_{LN}(\mathcal{F}_{MA}(\boldsymbol{W}^Q\boldsymbol{t},\boldsymbol{W}^K\boldsymbol{t},\boldsymbol{W}^V\boldsymbol{t}) + \boldsymbol{t}), \label{eq17}\\
  \boldsymbol{p} &= \mathcal{F}_{LN}(\mathcal{F}_{FF}(\boldsymbol{z}_{self})+ \boldsymbol{z}_{self}), \label{eq18}
\end{align}
where $\boldsymbol{z}_{self}$ is the output of the multi-head self-attention processed by layer normalization and residual connection with embeddings $\boldsymbol{t}$ as input, $\boldsymbol{W}^Q$, $\boldsymbol{W}^K$ and $\boldsymbol{W}^V$ are the weights parameters, \hlbreakable{$\boldsymbol{p}$ denotes the semantic encoded information in (\ref{eq1}) and also the output of the Transformer layer with a size determined by sentence length and output units of the semantic encoder.}

For the channel encoder and decoder, fully connected dense layers are used to encode and recover the information from the corrupted channel condition, which can be expressed as,
\begin{gather}
  \mathcal{F}_{FC}(\boldsymbol{X}) = \rho(\boldsymbol{W}\boldsymbol{X}+\boldsymbol{b}), \label{eq19}\\
  \boldsymbol{x} = \mathcal{F}_{FC}(\boldsymbol{p},\boldsymbol{\beta}),\label{eq20}\\
  \boldsymbol{\hat{p}} = \mathcal{F}_{FC}(\boldsymbol{\hat{x}},\boldsymbol{\gamma})), \label{eq21}
\end{gather}
where $\mathcal{F}_{FC}(\cdot)$ denote the fully connected layer with input $\boldsymbol{X}$ and parameters $\boldsymbol{W}$ and $\boldsymbol{b}$, $\rho$ is the activation function, $\boldsymbol{x}$ denotes the channel encoded information in (\ref{eq2}) with semantic encoded information $\boldsymbol{p}$ as input and $\boldsymbol{\beta}$ as trainable parameters, $\boldsymbol{\hat{p}}$ denotes the channel decoded information with recovered symbols $\boldsymbol{\hat{x}}$ at receiver as input and $\boldsymbol{\gamma}$ as trainable parameters. 

Different from the semantic encoder, where the attention is calculated instantaneous for the entire sequence, the semantic decoder estimates the sequence by iteratively predicting each word sequentially using the previous estimate as input. Therefore, an additional self-attention for the predicted words of each iteration is required for the semantic decoder. The Transformer layer of the semantic decoder can be expressed as
\begin{gather}
  \boldsymbol{z}' = \mathcal{F}_{LN}(\mathcal{F}_{MA}(\boldsymbol{W}'^Q\boldsymbol{\hat{t}}',\boldsymbol{W}'^K\boldsymbol{\hat{t}}',\boldsymbol{W}'^V\boldsymbol{\hat{t}}')+ \boldsymbol{\hat{t}}'), \label{eq22}\\
  \boldsymbol{z}_{cross} = \mathcal{F}_{LN}(\mathcal{F}_{MA}(\boldsymbol{W}^Q\boldsymbol{z}',\boldsymbol{W}^K\hat{\boldsymbol{p}},\boldsymbol{W}^V\hat{\boldsymbol{p}}) + \boldsymbol{z}'),\label{eq23}\\
  \boldsymbol{\hat{t}} = \mathcal{F}_{LN}(\mathcal{F}_{FF}(\boldsymbol{z}_{cross})+ \boldsymbol{z}_{cross}),\label{eq24}
\end{gather}
where $\boldsymbol{z}'$ denotes the multi-head self-attention with the predicted embeddings $\boldsymbol{\hat{t}}'$ as input, $\boldsymbol{\hat{t}}' \subseteq \boldsymbol{\hat{t}}$ denotes the results of each iteration to estimate the sequence, $\boldsymbol{W}'^Q$, $\boldsymbol{W}'^K$ and $\boldsymbol{W}'^V$ denote the parameters for the self-attention of the previous predictions, $\boldsymbol{z}_{cross}$ is the multi-head cross-attention with attention of the previous predictions $\boldsymbol{z}'$ and channel decoded information $\hat{\boldsymbol{p}}$ as input, $\boldsymbol{W}^Q$, $\boldsymbol{W}^K$ and $\boldsymbol{W}^V$ denote the parameters for the multi-head cross-attention, $\boldsymbol{\hat{p}}$ is the output of the channel decoder in (\ref{eq7}), $\boldsymbol{\hat{t}}$ is the output of the semantic decoder in (\ref{eq8}).

In training, we use the masked sequence embedding $\boldsymbol{\hat{t}}_{masked}$ as input to predict the masked word instead of using the output of the previous predictions $\boldsymbol{\hat{t}}'$ to predict the next word. The attention mechanism can learn the contextual information around the masked word during training. This could speed up the training and address the problems that the model cannot make reliable predictions when under-trained. During the deployment and testing, we always use the predictions of the previous iterations to estimate the next. Note that the semantic encoder-decoder and the channel encoder-decoder can have multiple layers, which are iterated by using the output of the current layer as the input for the next.
\subsection{KD-based system model}
\begin{figure*}[!ht]
  \centering
  \includegraphics[width=6.4in]{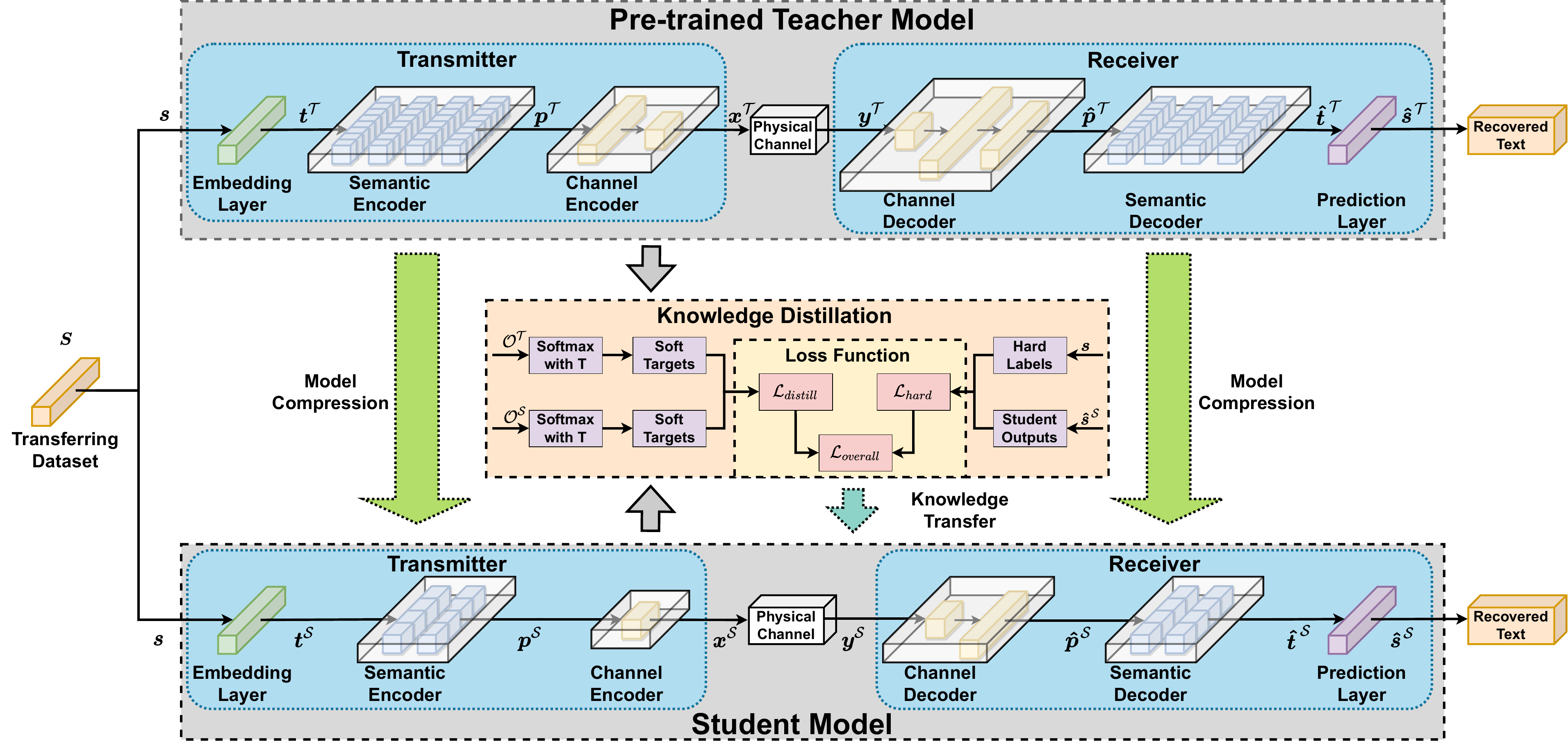}
  \caption{The structure of the KD-based SemCom system.}
  \label{kd_system}
\end{figure*}
As shown in Fig. \ref{kd_system}, we propose a tailored KD algorithm for the SemCom system. In KD, the teacher is required to provide soft targets as knowledge to train the student model. The soft targets are the probability distribution obtained by applying the softmax function to the output of the model. By adding the temperature parameter $T$ in the softmax, it could control the level of uncertainty in the output probabilities. By raising the temperature, the soft targets become more diffuse with less emphasis on the most probable class, which can help prevent the student model from overfitting to the training data and encourage it to learn more generalizable features \cite{44873}. The softmax function with temperature $T$ can be described by
\begin{equation}
  Q(z_i; T) = \frac{exp(z_i/T)}{\sum_{j} exp(z_j/T)}, \label{eq25}
\end{equation}
where $z_i$ is the $i$-th output of the model which can be the final prediction or the intermediate feature representation, and $T$ denotes the temperature parameter. To transfer knowledge from teacher to student, instead of only considering the final output logits, we consider the intermediate outputs in the SemCom system including the encoded semantic information $\boldsymbol{p}^\mathcal{T}$, $\boldsymbol{p}^\mathcal{S}$, the encoded channel information $\boldsymbol{x}^\mathcal{T}$, $\boldsymbol{x}^\mathcal{S}$, the decoded channel information $\boldsymbol{\hat{p}}^\mathcal{T}$, $\boldsymbol{\hat{p}}^\mathcal{S}$, the decoded semantic information $\boldsymbol{\hat{t}}^\mathcal{T}$, $\boldsymbol{\hat{t}}^\mathcal{S}$ and the final predictions $\boldsymbol{\hat{s}}^\mathcal{T}$, $\boldsymbol{\hat{s}}^\mathcal{S}$. \hlbreakable{In this way, teachers' intermediate outputs can be used as additional supervisory data to guide student training and as a comparative benchmark to implicitly explain students' intermediate outputs in the SemCom system. Taking semantic encoded information as an example, tests show that a closer distribution to the teacher model could enhance the capability of the student model.}

The total distillation loss is computed by  
\begin{equation}
  \begin{split}
    &\mathcal{L}_{total\_distill} = \sum_{(\mathcal{O}^\mathcal{S}, \mathcal{O}^\mathcal{T}) \in \mathbb{O}} \mathcal{L}_{distill}(\mathcal{O}^\mathcal{S}, \mathcal{O}^\mathcal{T})\\
   &= \sum_{(\mathcal{O}^\mathcal{S}, \mathcal{O}^\mathcal{T}) \in \mathbb{O}} \mathbb{E}\Big\{   \eta_{\mathcal{O}^\mathcal{S}, \mathcal{O}^\mathcal{T}}T^2 \mathcal{D}_{KL}[Q(\mathcal{O}^{\mathcal{S}};T)|| Q(\mathcal{O}^{\mathcal{T}};T)]\Big\},
  \end{split}
\end{equation} 
\begin{equation}
\mathbb{O} \subseteq  \Big\{(\boldsymbol{p}^\mathcal{S},\boldsymbol{p}^\mathcal{T}),(\boldsymbol{x}^\mathcal{S},\boldsymbol{x}^\mathcal{T}),(\boldsymbol{\hat{p}}^\mathcal{S},\boldsymbol{\hat{p}}^\mathcal{T}),(\boldsymbol{\hat{t}}^\mathcal{S},\boldsymbol{\hat{t}}^\mathcal{T}),(\boldsymbol{\hat{s}}^\mathcal{S},\boldsymbol{\hat{s}}^\mathcal{T}) \Big\}, 
\end{equation}  
where $\mathcal{L}_{distill}(\cdot)$ is the corresponding distillation loss of each output from the teacher and the student; $\mathcal{O}^\mathcal{T}$ denotes the output from the teacher serving as the reference to the output from the student $\mathcal{O}^\mathcal{S}$; $\mathcal{O}^\mathcal{T}$ and $\mathcal{O}^\mathcal{S}$ are the subset of the distillation information set $\mathbb{O}$; $\eta_{\mathcal{O}^\mathcal{S},\mathcal{O}^\mathcal{T}}$ is the weight parameter for each distillation loss; $\mathcal{D}_{KL}(\cdot)$ denotes the Kullback-Leibler (KL) divergence \cite{Kullback51klDivergence}. KL divergence can compare two probability distributions with different scales, which provides a way to measure how much the student distribution deviates from the teacher distribution. Minimizing this difference enables the student to learn from the teacher and reproduce the teacher's probability distribution. Additionally, KL divergence allows the temperature parameter $T$ to be adjusted, which improves the flexibility of the training. To obtain the overall training loss for the student model, we combine the cross entropy loss with the hard labels and the sum of the distillation loss, which can be represented as, 
\begin{equation}\label{loss_overall}
  \begin{split}
    &\mathcal{L}_{overall} = (1-\sum_{(\mathcal{O}^\mathcal{T}, \mathcal{O}^\mathcal{S}) \in \mathbb{O}} \eta_{\mathcal{O}^\mathcal{T},\mathcal{O}^\mathcal{S}}) \mathcal{L}_{CE}(\boldsymbol{s}^\mathcal{S},\boldsymbol{\hat{s}}^\mathcal{S}) \\
    +&\sum_{(\mathcal{O}^\mathcal{T}, \mathcal{O}^\mathcal{S}) \in \mathbb{O}} \mathbb{E} \Big\{ \eta_{\mathcal{O}^\mathcal{T},\mathcal{O}^\mathcal{S}} T^2 \mathcal{D}_{KL}(Q(\mathcal{O}^{\mathcal{S}};T)|| Q(\mathcal{O}^{\mathcal{T}};T))\Big\}.
  \end{split}
\end{equation}
Then, the gradient of the overall loss is calculated and back-propagated to update the parameters. Therefore, the error between the teacher's output probability distributions and student output distributions can be minimized. In other words, the optimization for the student can be guided by the teacher model, which is equivalent to matching the corresponding outputs of each part from the teacher model to the student model \cite{44873}.

\subsection{Model compression}
The proposed SemCom system adopts the Transformer structure as the semantic encoder-decoder and dense layers as the channel encoder-decoder. Unlike the conventional learning-based system that is treated as a black box, the training of the proposed KD-based SemCom system can be divided into several small black boxes based on the distilled knowledge. Each black box can converge the corresponding outputs from the pre-trained teacher model. This process can potentially have more control on the optimization process of the student model. Therefore, we can conduct the model compression by reducing the size of each component in the student model while converging the outputs, as shown in Fig. \ref{kd_system}. However, simply reducing the number of Transformers in the semantic encoder-decoder could weaken the robustness of the model. To alleviate this nagative effect, the student learns from the teacher by mimicking the teacher's outputs to improve generalization ability since the teacher model is over-parameterized and pre-trained with extensive data. This is achieved by minimizing $\mathcal{L}_{overall}$ in equation (\ref{loss_overall}). Similarly, the number of dense layers in the channel encoder-decoder can also be reduced for model compression. Although this might affect the information recovery performance, the knowledge from the teacher is transferred to compensate for the impact on the performance. 

Also, inspired by the work in \cite{9252948} using network quantization for the SemCom system, we propose to use post training dynamic quantization to further compress our model after reducing the number of layers and parameters via KD. Dynamic quantization converts the float representation of the weights to the reduced integer representation, which essentially saves the model size and computational complexity. The weights quantization can be expressed as,
\begin{equation}
  \boldsymbol{X}_q = round(\varphi \boldsymbol{X}_{float}-\omega), \label{quan}
\end{equation}
where $\boldsymbol{X}_q$ is the quantized output, $\boldsymbol{X}_{float}$ is the float input, $\varphi$ is the scale parameter and $\omega$ is the zero point. Note that, overly decreasing the size of the mode could cause the model to diverge and lose generalization ability. The trade-off between performance and the size of the model will be discussed to answer Question 3). 

\subsection{Training}
\begin{algorithm}[ht]
  \caption{Data generation}
  \hlbreakable{\begin{algorithmic}[1]
    \REQUIRE: Dataset $\boldsymbol{S}$, number of interfering users $N$, SemCom model $\mathcal{SE}(\cdot)$, $\mathcal{CE}(\cdot)$, $\mathcal{CD}(\cdot)$, $\mathcal{SD}(\cdot)$ and $\mathcal{P}_{pred}(\cdot)$ with parameters $\Theta=\{\boldsymbol{\alpha}$,$\boldsymbol{\beta}$,$\boldsymbol{\gamma}$,$\boldsymbol{\delta}$, $\boldsymbol{w}_{pred}$,$\boldsymbol{b}_{pred}\}$
    \STATE   $\boldsymbol{s} \gets BatchDataset(\boldsymbol{S})$.
    \STATE   $\boldsymbol{t} \gets Embedding(\boldsymbol{s})$.
    \STATE Compute the output of semantic encoder by (\ref{eq17}) and (\ref{eq18}), $\boldsymbol{p} \gets \mathcal{SE}(\boldsymbol{t},\boldsymbol{\alpha})$.
    \STATE Compute the output of channel encoder by (\ref{eq19}) and (\ref{eq20}), $\boldsymbol{x} \gets \mathcal{CE}(\boldsymbol{p},\boldsymbol{\beta})$.
    \IF{ Train the teacher model }
        \STATE Transmit $\boldsymbol{x}$ over the physical channel with MU interference $\boldsymbol{i}^\mathcal{I}_1,...,\boldsymbol{i}^\mathcal{I}_N$ in (3).
    \ELSIF{ Train the student model }
        \STATE Transmit $\boldsymbol{x}$ over the physical channel with no MU interference.
    \ENDIF
    \STATE  Receive $\boldsymbol{y}$.
    \STATE  Compute the output of channel decoder by (\ref{eq19}) and (\ref{eq21}), $\boldsymbol{\hat{p}} \gets \mathcal{CD}(\boldsymbol{y},\boldsymbol{\gamma})$.
    \STATE  Compute the output of semantic decoder by (\ref{eq22}), (\ref{eq23}) and (\ref{eq24}), $\boldsymbol{\hat{t}} \gets \mathcal{SD}(\boldsymbol{\hat{p}},\boldsymbol{\delta})$.
    \STATE  Compute the predicted results by (\ref{eq9}), $\boldsymbol{\hat{s}} \gets \mathcal{P}_{pred}(\boldsymbol{\hat{t}},\boldsymbol{w}_{pred},\boldsymbol{b}_{pred})$.
    \ENSURE $\boldsymbol{s}$, $\boldsymbol{p}$, $\boldsymbol{x}$, $\hat{\boldsymbol{p}}$, $\hat{\boldsymbol{t}}$ and $\hat{\boldsymbol{s}}$. 
    \end{algorithmic}}
\end{algorithm}
To train the KD-based SemCom system, the combined training process is demonstrated in Algorithm 4. It can be divided into three phases: training the teacher model, training the student model and applying post-training quantization. Algorithm 1 demonstrates the feed forward process to generate the outputs for teacher and student models. \hlbreakable{Since the proposed SemCom system adopts a DL-based E2E transceiver design, the feed-forward process is implemented as the complete process of encoding the information at the transmitter and recovering it at the receiver. In order to generate inputs for the semantic encoder, the dataset is partitioned into batches for parallel processing during training, and each word is then transformed into word embeddings for use as inputs. The semantic encoder consists of multiple Transformer layers with the multi-head self-attention in (\ref{eq17}) and (\ref{eq18}), while the semantic decoder additionally computes the multi-head cross-attention to account for previous predictions, denoted in (\ref{eq22}), (\ref{eq23}) and (\ref{eq24}). For channel encoder and decoder, it adopts multiple fully connected dense layers in (\ref{eq19}).}
\begin{algorithm}[t]
  \caption{Training algorithm of the teacher model}
  \begin{algorithmic}[1]
    \REQUIRE Training dataset $\boldsymbol{S}^\mathcal{T}$, MU interfering model $\mathcal{SE}^\mathcal{I}(\cdot)$, $\mathcal{CE}^\mathcal{I}(\cdot)$, number of epochs $E$, Teacher model $\mathcal{SE}^\mathcal{T}(\cdot)$, $\mathcal{CE}^\mathcal{T}(\cdot)$, $\mathcal{CD}^\mathcal{T}(\cdot)$, $\mathcal{SD}^\mathcal{T}(\cdot)$ and $\mathcal{P}_{pred}^\mathcal{T}(\cdot)$ with parameters $\Theta^\mathcal{T}=\{\boldsymbol{\alpha}^\mathcal{T}$,$\boldsymbol{\beta}^\mathcal{T}$,$\boldsymbol{\gamma}^\mathcal{T}$,$\boldsymbol{\delta}^\mathcal{T}$, $\boldsymbol{w}_{pred}^\mathcal{T}$,$\boldsymbol{b}_{pred}^\mathcal{T}\}$.
    \STATE Initialize parameters $\Theta^\mathcal{T}$.
    \FOR{$e = 1$ to $E$}
    \STATE Perform forward propagation to compute output $\boldsymbol{\hat{s}}^\mathcal{T}$.
    \STATE Compute cost $J(\Theta^\mathcal{T})$ using loss function $\mathcal{L}_{hard}$ in (\ref{eq10}).
    \STATE  Perform backward propagation to compute gradients $\frac{\partial J}{\partial \Theta^\mathcal{T}}$.
    \STATE  Update parameters $\Theta^\mathcal{T}$ using stochastic gradient descent.
    \ENDFOR
    \ENSURE Trained $\mathcal{SE}^\mathcal{T}(\cdot)$, $\mathcal{CE}^\mathcal{T}(\cdot)$, $\mathcal{CD}^\mathcal{T}(\cdot)$, $\mathcal{SD}^\mathcal{T}(\cdot)$ and $\mathcal{P}_{pred}^\mathcal{T}(\cdot)$ with parameters $\Theta^\mathcal{T}$.
  \end{algorithmic}
\end{algorithm}
As the teacher model is pre-trained, we assume that the teacher model has sufficient computational resources and large amount of training data. Also, we simulate different SNRs on the transmitted signals. With these samples in the training dataset, the teacher can have robust performance in the scenarios across different SNRs, with or without interference. The training algorithm for the teacher model is demonstrated in Algorithm 2. Firstly, the feed forward process is applied to generate the teacher's outputs, which introduces the interference signals from the interfering semantic transmitters. Afterwards, the cost is computed by calculating cross-entropy loss and propagating back to compute gradients. Then, the stochastic gradient descent is used to update the parameters in the semantic transceiver.
\begin{algorithm}[ht]
  \caption{Training algorithm of the student model}
  \begin{algorithmic}[1]
    \REQUIRE Training dataset $\boldsymbol{S}^\mathcal{S}$, number of epochs $E$, pre-trained teacher model $\mathcal{SE}^\mathcal{T}(\cdot)$, $\mathcal{CE}^\mathcal{T}(\cdot)$, $\mathcal{CD}^\mathcal{T}(\cdot)$, $\mathcal{SD}^\mathcal{T}(\cdot)$ and $\mathcal{P}_{pred}^\mathcal{T}(\cdot)$, student model $\mathcal{SE}^\mathcal{S}(\cdot)$, $\mathcal{CE}^\mathcal{S}(\cdot)$, $\mathcal{CD}^\mathcal{S}(\cdot)$, $\mathcal{SD}^\mathcal{S}(\cdot)$ and $\mathcal{P}_{pred}^\mathcal{S}(\cdot)$ with parameters $\Theta^\mathcal{S}=\{\boldsymbol{\alpha}^\mathcal{S}$,$\boldsymbol{\beta}^\mathcal{S}$,$\boldsymbol{\gamma}^\mathcal{S}$,$\boldsymbol{\delta}^\mathcal{S}$, $\boldsymbol{w}_{pred}^\mathcal{S}$,$\boldsymbol{b}_{pred}^\mathcal{S}\}$.
    \STATE Initialize parameters $\Theta^\mathcal{S}$, load pretrained teacher $\mathcal{SE}^\mathcal{T}(\cdot)$, $\mathcal{CE}^\mathcal{T}(\cdot)$, $\mathcal{CD}^\mathcal{T}(\cdot)$, $\mathcal{SD}^\mathcal{T}(\cdot)$ and $\mathcal{P}_{pred}^\mathcal{T}(\cdot)$
    \FOR{$e = 1$ to $E$}
    \STATE Compute the outputs of pretrained teacher $\boldsymbol{p}^\mathcal{T}$, $\boldsymbol{x}^\mathcal{T}$, $\hat{\boldsymbol{p}}^\mathcal{T}$, $\hat{\boldsymbol{t}}^\mathcal{T}$ and $\hat{\boldsymbol{s}}^\mathcal{T}$.
    \STATE Perform forward propagation to compute the student outputs $\boldsymbol{p}^\mathcal{S}$, $\boldsymbol{x}^\mathcal{S}$, $\hat{\boldsymbol{p}}^\mathcal{S}$, $\hat{\boldsymbol{t}}^\mathcal{S}$ and $\hat{\boldsymbol{s}}^\mathcal{S}$.
    \STATE Compute cost $J(\Theta^\mathcal{S})$ using loss function $\mathcal{L}_{overall}$ in (\ref{loss_overall}).
    \STATE  Perform backward propagation to compute gradients $\frac{\partial J}{\partial \Theta^\mathcal{S}}$.
    \STATE  Update parameters $\Theta^\mathcal{S}$ using stochastic gradient descent.
    \ENDFOR
    \ENSURE Trained $\mathcal{SE}^\mathcal{S}(\cdot)$, $\mathcal{CE}^\mathcal{S}(\cdot)$, $\mathcal{CD}^\mathcal{S}(\cdot)$, $\mathcal{SD}^\mathcal{S}(\cdot)$ and $\mathcal{P}_{pred}^\mathcal{S}(\cdot)$ with parameters $\Theta^\mathcal{S}$.
  \end{algorithmic}
\end{algorithm}
\begin{algorithm}[ht]
  \caption{Combined training algorithm for the proposed KD-based SemCom }
  \begin{algorithmic}[1]
    \REQUIRE Training dataset $\boldsymbol{S}^\mathcal{T}$, transfer dataset $\boldsymbol{S}^\mathcal{S}$.
    \STATE   Train the teacher model $\gets \boldsymbol{S}^\mathcal{T}$.
    \STATE   Train the student model $\gets \boldsymbol{S}^\mathcal{S}$, pretrained teacher model $\mathcal{SE}^\mathcal{T}(\cdot)$, $\mathcal{CE}^\mathcal{T}(\cdot)$, $\mathcal{CD}^\mathcal{T}(\cdot)$, $\mathcal{SD}^\mathcal{T}(\cdot)$ and $\mathcal{P}_{pred}^\mathcal{T}(\cdot)$.
    \STATE  Quantize the student model by (\ref{quan}).
    \ENSURE: Trained and quantized $\mathcal{SE}^\mathcal{S}(\cdot)$, $\mathcal{CE}^\mathcal{S}(\cdot)$, $\mathcal{CD}^\mathcal{S}(\cdot)$, $\mathcal{SD}^\mathcal{S}(\cdot)$ and $\mathcal{P}_{pred}^\mathcal{S}(\cdot)$.
  \end{algorithmic}
\end{algorithm}

After obtaining the trained teacher model, we train the student model in the second phase, which is demonstrated in Algorithm 3. We use the transfer training dataset $\boldsymbol{S}^\mathcal{S}$, which contains limited samples to simulate the scenario when there are not enough training data. Unlike the training dataset for the teacher model, we do not add MU interference in this training data and simulate a limited range of SNRs. Considering these constraints, we apply the feed forward propagation in Algorithm 1 to obtain the outputs of the students and the teacher from semantic encoder, channel encoder, channel decoder, semantic decoder, and prediction layer, respectively. Then, we apply the KD algorithm by computing the overall loss $\mathcal{L}_{overall}$, which considers the distilled loss $\mathcal{L}_{total\_distill}$ between the student and the teacher and the cross-entropy loss between the predicted sentence and the ground truth. Then, the student model's parameters are optimized by stochastic gradient descent after backpropagation. In the third phase, we apply post-training quantization to the weights and activation function so that the model size can be further compressed.

The overall loss function using cross-entropy and KL divergence could have multiple local minima due to their asymmetric inputs. Besides, the semantic encoder-decoder and channel encoder-decoder consists of millions of parameters with several hidden layers and non-linear activation function, which also causes multiple local minima. Therefore, the optimization problem using KD is considered non-convex. To alleviate the non-convexity and approximate the global minimum, we adopt Adam optimizer \cite{kingma2014adam}, layer normalization and dropout layer in the network to avoid local minima and improve convergence.

\section{Numerical results and discussion} \label{sec4}
In this section, we examine the performance of the proposed KD-based SemCom system in Rayleigh fading channels with MU interference. We assume perfect CSI for the desired user and various SNR regime is simulated in the experiment. 
\subsection{Simulation setting}
There are several types of knowledge from the teacher model, and increasing distilled knowledge for the student model can increase the difficulty of the training by introducing more hyperparameters. Therefore, we propose four types of student models to evaluate the performance for different distilled knowledge and model compression. These four models will be trained using the limited dataset and tested in the SemCom system in the presence of co-channel interference to analyze the generalization ability on unseen data, which could answer Questions 1) and 2). 

\begin{table*}[!t]
\caption{The setting of the teacher and student models for the SemCom system}
\label{kd-student}
\renewcommand{\arraystretch}{1}
\centering
\resizebox{4.5in}{!}{%
\begin{tabular}{|c|ccccc|}
\hline
 &
  \multicolumn{1}{c|}{Teacher} &
  \multicolumn{1}{c|}{Student 1} &
  \multicolumn{1}{c|}{Student 2} &
  \multicolumn{1}{c|}{Student 3} &
  Student 4 \\ \hline
\begin{tabular}[c]{@{}c@{}}Semantic\\ encoder\end{tabular} &
  \multicolumn{2}{c|}{\begin{tabular}[c]{@{}c@{}}4 x Transformer layers\\ 8 heads 128 units\end{tabular}} &
  \multicolumn{3}{c|}{\begin{tabular}[c]{@{}c@{}}2 x Transformer layers\\ 8 heads 128 units\end{tabular}} \\ \hline
\begin{tabular}[c]{@{}c@{}}Channel\\ encoder\end{tabular} &
  \multicolumn{3}{c|}{\begin{tabular}[c]{@{}c@{}}2 x Dense layers\\ 256, 16 units\end{tabular}} &
  \multicolumn{2}{c|}{\begin{tabular}[c]{@{}c@{}}1 x Dense layer\\ 16 units\end{tabular}} \\ \hline
\begin{tabular}[c]{@{}c@{}}Channel\\ decoder\end{tabular} &
  \multicolumn{3}{c|}{\begin{tabular}[c]{@{}c@{}}3 x Dense layers\\ 128, 512, 128 units\end{tabular}} &
  \multicolumn{2}{c|}{\begin{tabular}[c]{@{}c@{}}2 x Dense layers\\ 128, 128 units\end{tabular}} \\ \hline
\begin{tabular}[c]{@{}c@{}}Semantic\\ decoder\end{tabular} &
  \multicolumn{2}{c|}{\begin{tabular}[c]{@{}c@{}}4 x Transformer layers\\ 8 heads 128 units\end{tabular}} &
  \multicolumn{3}{c|}{\begin{tabular}[c]{@{}c@{}}2 x Transformer layers\\ 8 heads 128 units\end{tabular}} \\ \hline
\begin{tabular}[c]{@{}c@{}}Prediction\\ layer\end{tabular} &
  \multicolumn{5}{c|}{\begin{tabular}[c]{@{}c@{}}1 x Dense layer\\ dictionary units\end{tabular}} \\ \hline
\begin{tabular}[c]{@{}c@{}}Distilled\\ knowledge\end{tabular} &
  \multicolumn{1}{c|}{-} &
  \multicolumn{2}{c|}{\begin{tabular}[c]{@{}c@{}}$(\boldsymbol{x}^\mathcal{S},\boldsymbol{x}^\mathcal{T})$,$(\boldsymbol{\hat{t}}^\mathcal{S},\boldsymbol{\hat{t}}^\mathcal{T})$ \\and $(\boldsymbol{\hat{s}}^\mathcal{S},\boldsymbol{\hat{s}}^\mathcal{T})$\end{tabular}} &
  \multicolumn{2}{c|}{\begin{tabular}[c]{@{}c@{}}$(\boldsymbol{p}^\mathcal{S},\boldsymbol{p}^\mathcal{T})$, $(\boldsymbol{x}^\mathcal{S},\boldsymbol{x}^\mathcal{T})$, $(\boldsymbol{\hat{p}}^\mathcal{S},\boldsymbol{\hat{p}}^\mathcal{T})$, \\ $(\boldsymbol{\hat{t}}^\mathcal{S},\boldsymbol{\hat{t}}^\mathcal{T})$ and $(\boldsymbol{\hat{s}}^\mathcal{S},\boldsymbol{\hat{s}}^\mathcal{T})$\end{tabular}} \\ \hline
Quantization &
  \multicolumn{1}{c|}{-} &
  \multicolumn{1}{c|}{-} &
  \multicolumn{1}{c|}{-} &
  \multicolumn{1}{c|}{-} &
  \checkmark \\ \hline
\end{tabular}%
}
\end{table*}
The parameter settings for the teacher and the student models are shown in Table \ref{kd-student}. We adopt the same structure as the L-DeepSC \cite{9252948} for Teacher and Student 1, which has 4 layers of Transformer with 8 attention heads and 128 units as the semantic encoder and decoder. Additionally, 2 dense layers with 256 and 16 units are used for the channel encoder, and 3 dense layers with 128, 512, and 128 units are used for the channel decoder. Student 2 adopts 2 Transformer layers for the semantic encoder and decoder, while Student 3 and 4 further reduce the model size by using 1 dense layer for channel encoder and 2 dense layers for channel decoder. Finally, a dense layer is adopted as the prediction layer to output a vector with a dictionary size to represent the predicted word. All of the student models conduct KD from the teacher model. Students 1 and 2 consider the knowledge from the outputs of channel encoder, semantic encoder, and prediction layer, while Students 3 and 4 additionally consider the outputs from the semantic encoder and channel decoder. Moreover, post-training dynamic quantization is applied to Student 4. All learning-based SemCom systems require 8 symbols to represent one word since the output units of the channel encoder are set as 16, and it is converted to a two-dimensional vector as complex for transmission.

The performance of all baseline models is evaluated using the same number of transmitted symbols to guarantee a fair comparison. The settings of the baselines are described below:
\subsubsection{Baselines of learning-based SemCom systems}All baselines are trained without KD to benchmark the contribution of distilled knowledge.
\begin{itemize}
 \item DeepSC \cite{9398576}: We apply the same structure of DeepSC, which has 3 layers of Transformer with 8 attention heads as semantic encoder and decoder and 2 dense layers as channel encoder and decoder. 
\item Baseline 1: We adopt the same structure as L-DeepSC, Teacher and Student 1. 
\item Baseline 2: We adopt the same model structure as Student 2. 
\item Baseline 3: We adopt the same model structure as Student 3. 
\end{itemize}

\subsubsection{Conventional communications systems}We adopt 8 symbols to represent one word for different source coding methods by choosing appropriate code rate for low-density parity-check code (LDPC) and 16-QAM for modulation, which are the same as the learning-based SemCom system.
\begin{itemize}
      \item Huffman coding and LDPC: Huffman coding requires about 20 bits to represent a word, and we adopt 16-QAM for modulation and 5/8 as the code rate for LDPC.
      \item 5-Bit coding and LDPC: We adopt 16-QAM and 7/8 for the code rate of LDPC, since 5-bit uses about 28 bits to represent a word. Also, 5-bit coding is used to benchmark the performance without compression. 
  \end{itemize}

The source of the training dataset is the English dataset from the Europarl \cite{koehn-2005-europarl}, which contains over 2 million sentences and 53 million words. The sentences are randomly split into the training set and the testing set. As the teacher model is pre-trained, we conduct different experiments to randomly add extra interference and noise to formulate the training datasets for the teacher model to make it robust. To simulate the interference, we use randomly selected sources. Moreover, the random occurrences and delays of interference are simulated by applying a 90\% of occurrence rate and a maximum three-word (24 symbols) delay to the interfering signals. Different from the extensive training datasets for the teacher, the transferring dataset for the student model and the baselines is formulated with an interference-free channel and a limited regime of noise, which does not contain any interference samples. In this case, the student models and the baselines are trained with limited transferring datasets compared with the teacher model. Therefore, we can test the student and baseline models with unseen interference to evaluate the robustness and generalization ability under limited training data. The difference between the student models and the baselines is that the student models are conducted KD, whereas the baselines are not. 

To measure the performance of the models mentioned above, we adopt a bilingual evaluation understudy (BLEU) score to measure the difference between two sentences \cite{papineni-etal-2002-bleu}. However, it is difficult for BLEU to distinguish synonyms or polysemy. Thus, we also use sentence similarity \cite{9398576}, which adopts BERT \cite{DBLP:conf/naacl/DevlinCLT19} to map the sentences into semantic vector space and compare their semantic vectors. The simulation is performed on a computer with Intel(R) Xeon(R) CPU E5-2678 v3 @ 2.50GHz and NVIDIA GeForce GTX 1080 Ti.

\subsection{Performance without interference}
\begin{figure}[!h]
  \centering
  \includegraphics[width=3.2in]{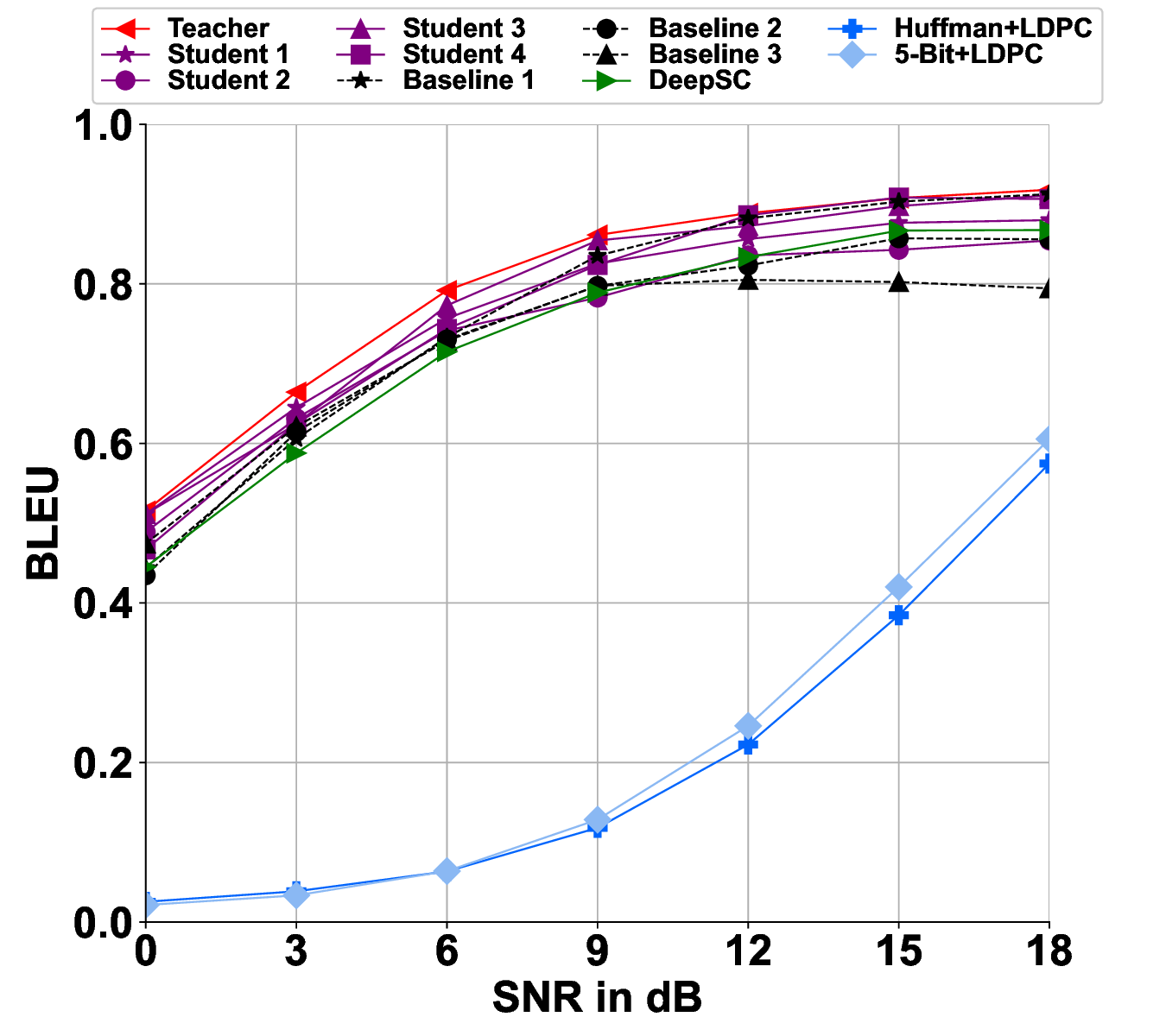}
  \caption{BLEU score of the SemCom with no interference.}
  \label{no_interference}
\end{figure}
In this section, we compare the performance without MU interference. The teacher model is trained using a dataset with SNR randomly changing from 10 dB to 15 dB, while others are trained by the transfer training dataset with a limited SNR ranging from 15 dB to 18 dB. Fig. \ref{no_interference} evaluates the BLEU performances for the teacher, the students, and the baselines. The BLEU accuracy of the teacher model ranges from about 52\% to over 90\% when the SNR increases from 0 dB to 18 dB. It outperforms other models because using the sufficient training set with a wider SNR range improves its robustness over different SNRs. With the transferred knowledge from the teacher model, all student models can perform better than the baselines without KD when the SNR is less than 9 dB. The reason is that the knowledge for the generalization in the low SNR regime is transferred from the teacher to the students, while the baselines can only learn through the hard label information. This also illustrates that, when the transfer dataset is limited, KD from the teacher can improve the robustness of the model. 

For conventional communications systems, the BLEU can reach about 60\% when SNR is 18 dB. The non-compression coding scheme, 5-Bit with LDPC, performs slightly better than Huffman and LDPC when SNR increases over 9 dB. However, there is still a significant performance gap compared with learning-based SemCom systems. Additionally, a slight performance decline for Baseline 3 can be observed when SNR increases over 12 dB due to overly simplifying the model without the distilled knowledge from the teacher. One of the advantages of KD is that the student model does not need to have access to a large amount of training data while the teacher can be trained offline anywhere. This allows for data isolation and privacy. Also, compared with conducting several experiments to train the student model for different values of SNRs and SIRs, directly distilling the knowledge from the teacher could save the experiment time.




\subsection{Performance with MU interference}
\begin{figure*}[!h]
  \centering
  \includegraphics[width=5.2in]{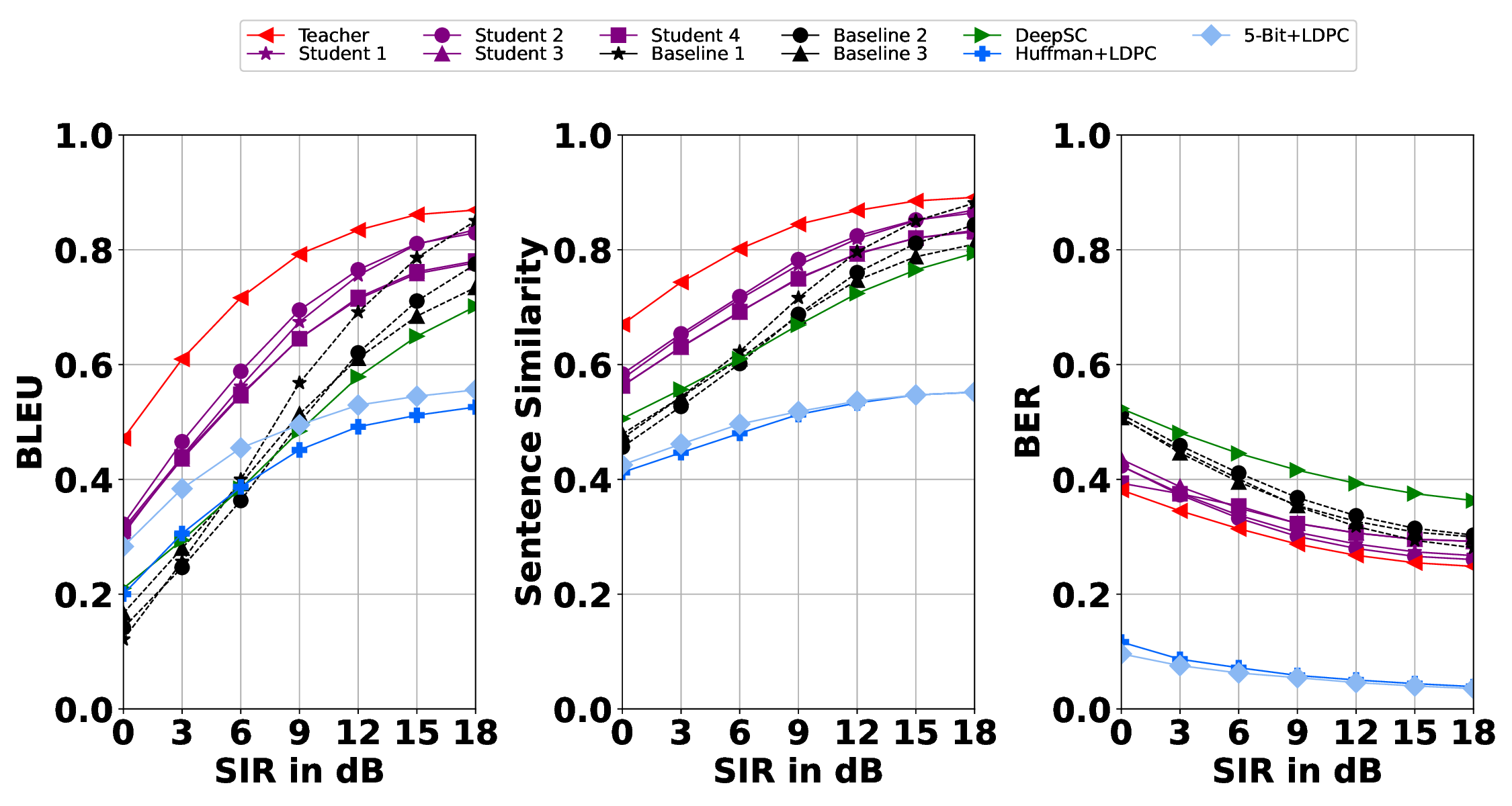}
  \caption{BLEU score and sentence similarity of the SemCom with one MU interference when the SNR is 18 dB.}
  \label{one_interference_sir}
\end{figure*}
\hlbreakable{To further investigate the effect of the proposed KD-based SemCom system, we evaluate the BLEU, sentence similarity and BER with one co-channel interference when SIR increases from 0 to 18 dB in Fig. \ref{one_interference_sir}. This is equivalent to increasing the number of interfering users while keeping the SIR for each user. The BER of the SemCom system is computed by converting the recovered words into bits using ASCII.} To evaluate the KD-based models with unseen data, the students and the baselines are trained with the transferring dataset without no MU interference samples, while they are tested in the presence of MU interference to show robustness. 

For the BLEU score, although the performance gap between the baselines and the student models narrows when SIR grows, the student models with distilled knowledge perform better than the baselines when SIR is less than 12 dB. For the conventional communications system, 5-bit and LDPC range from 30\% to over 50\% when SIR increases to 18 dB, which performs better than Huffman and LDPC. Nevertheless, the students with KD still greatly outperform the conventional communications systems regardless of the SIRs. The performance of sentence similarity in Fig. \ref{one_interference_sir} demonstrates the same trend as the BLEU scores. The sentence similarity of students ranges from about 60\% to over 80\% when SIR increases to 18 dB, which is better than the baselines and conventional communications systems. This shows that the models with KD have better word accuracy in BLEU scores and recover the sentences that are easier to understand. Moreover, the learning-based communications systems outperform the conventional communications system for all SIRs, indicating that people can better understand the text with the SemCom system. \hlbreakable{However, the conventional baselines do perform better than all the SemCom systems in terms of BER. This is attributed to the transformer-based encoder-decoder used by SemCom, which recovers the intended message through contextual reasoning on a word-by-word basis, instead of adhering strictly to the precise wording. SemCom sacrifices bit errors for the recovery of the meaning of words to save spectrum. In this regard, the semantic systems may have limited applications in services requiring precise bits, such as voice control. Despite this, the students with KD still outperform the learning-based baselines in BER, demonstrating the enhanced generalization to unseen interference by KD. }

\begin{figure}[!h]
  \centering
  \includegraphics[width=3.4in]{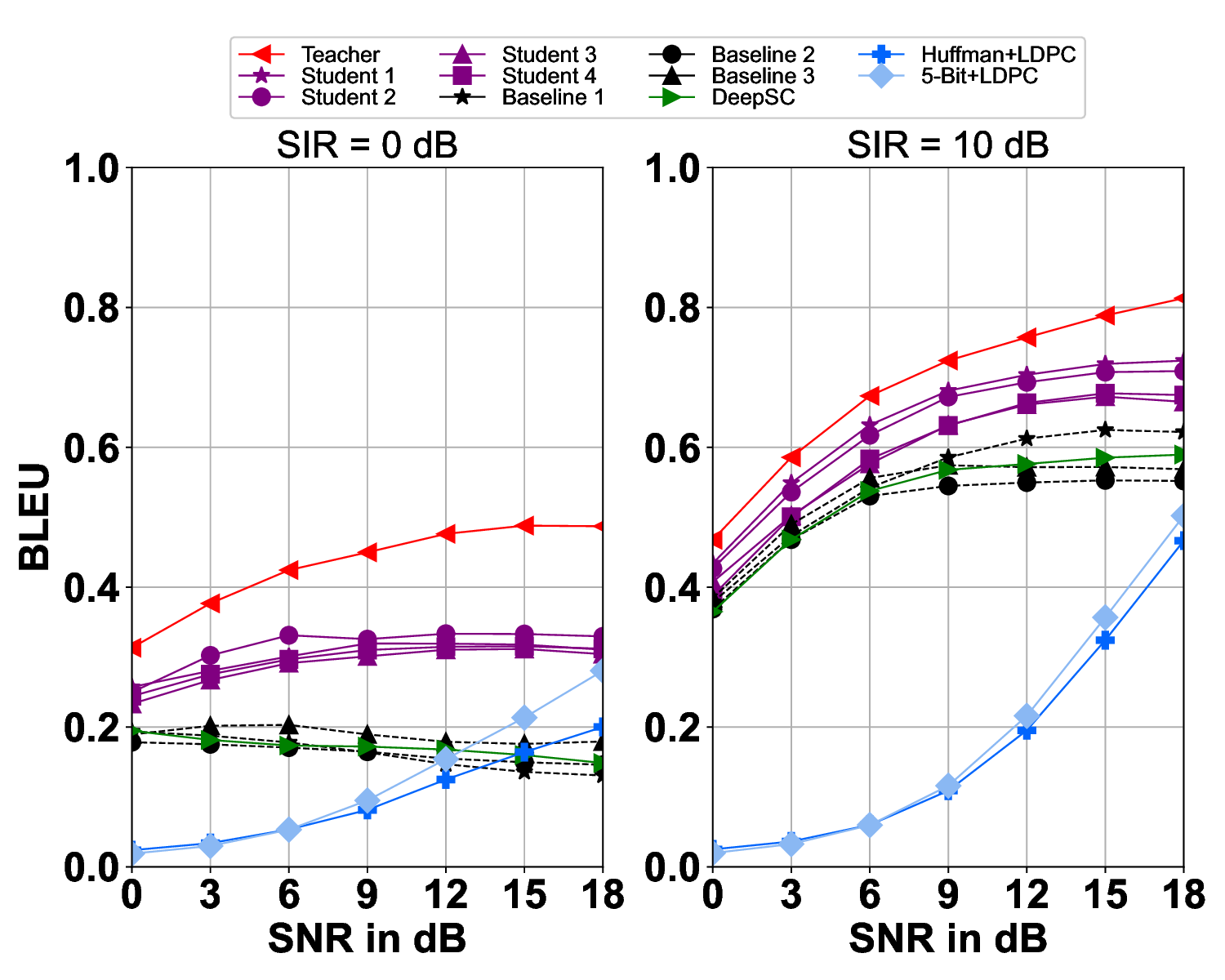}
  \caption{BLEU score of the SemCom with one MU interference.}
  \label{One_interference}
\end{figure}
Fig. \ref{One_interference} demonstrates the performances in the presence of one interfering user when SIR is 0 dB and 10 dB to simulate the strong and weak interference. When the physical channel is severely interfered, the BLEU accuracy of the baselines and DeepSC can barely reach 20\% because there is neither KD in the training nor interference samples in the training data. However, it still outperforms the conventional communications system with Huffman and 5-bit coding schemes when the SNR is less than 12 dB. The overall BLEU score of the teacher ranges from about 30\% to 50\% as the SNR increases from 0 to 18 dB due to its powerful generalization ability. On the other hand, the BLEU accuracy of the student models with distilled knowledge ranges from 25\% to 30\% when SNR increases from 0 to 18 dB, which outperforms the baselines and DeepSC by over 10\%. This shows that distilled knowledge can improve the model performance of generalizing on unseen data. Moreover, Student 2 and Student 1 have better performances than other students because overly compressing the size of the model can result in performance degradation. 

When the SIR is 10 dB, the BLEU score of the teacher ranges from about 47\% to over 80\% when the SNR is from 0 to 18 dB. Again, the student models perform better than the baselines and DeepSC from 0 to 18 dB. Also, Student 2 and Student 1 perform slightly better than Student 3 and Student 4. Student 2 only compresses the semantic encoder and decoder, while Student 3 and Student 4 conduct model compression on both semantic encoder-decoder and channel encoder-decoder. Additionally, Student 3 and Student 4 have similar performances, which illustrates that the post-training dynamic quantization can barely have any effect on the BLEU score performance. 
\begin{figure}[!h]
  \centering
  \includegraphics[width=3.4in]{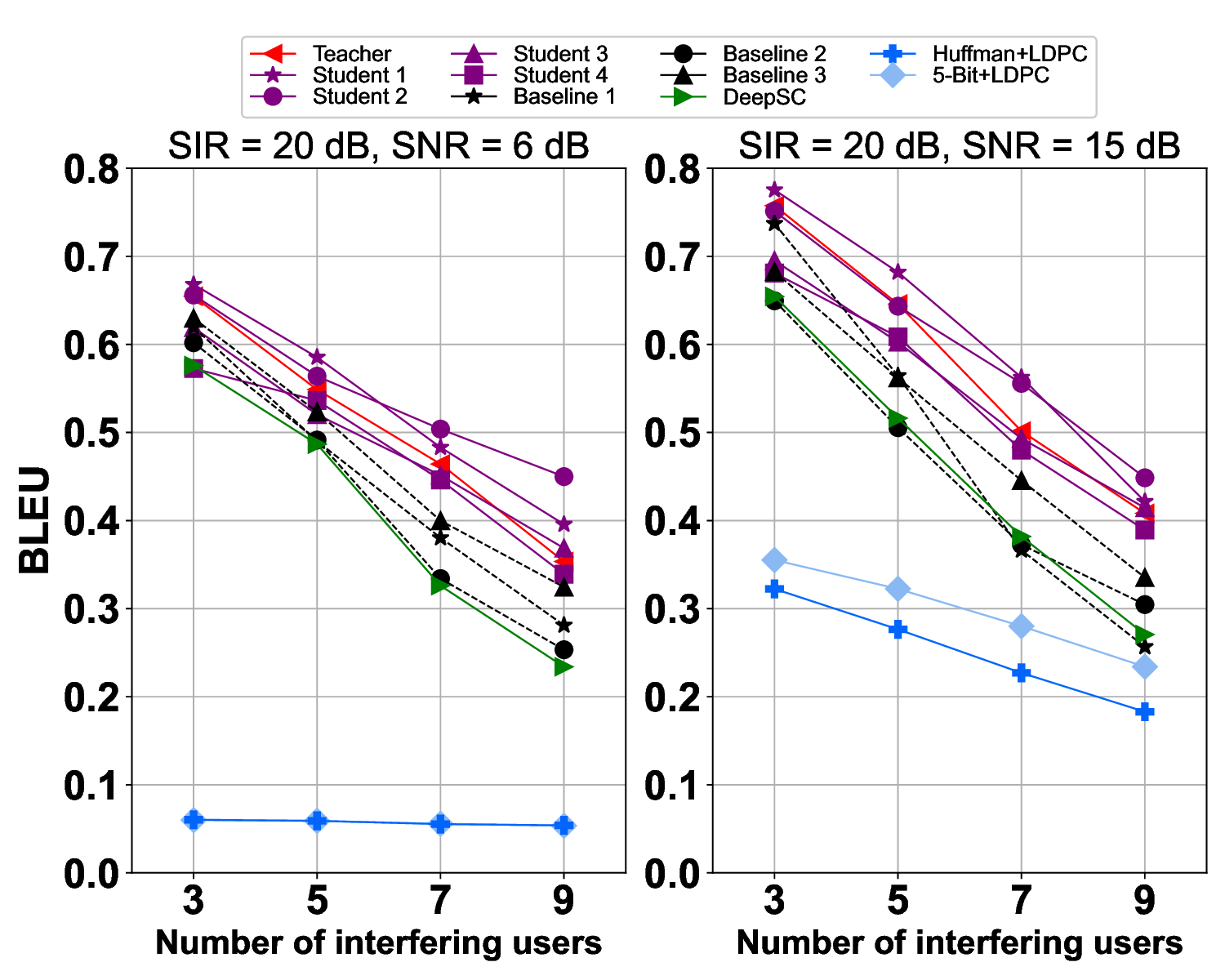}
  \caption{BLEU score of the SemCom with multiple interference.}
  \label{number_interference}
\end{figure}

To test the performance when there are more interfering users, Fig. \ref{number_interference} demonstrates the BLEU performance with multiple interference when SIR is 20 dB. When SNR is 6 dB, Student 1 and Student 2 outperform the teacher, as the number of interference increases from 3 to 9. This demonstrates that distilled knowledge can provide the generalization ability for Students 1 and 2, and direct the student models in optimization to outperform their teacher. Moreover, this compresses the model and improves the performance simultaneously. Additionally, all student models can generally outperform the baselines and DeepSC, except that Student 4 has a relatively poor accuracy when the number of interference is 3. For conventional communications systems, the BLEU of the Huffman and 5-bit coding with LDPC remains less than 10\%. When the SNR is 15 dB, the BLEU score of the teacher ranges from about 77\% to 41\% as the number of interfering users increases from 3 to 9. The 5-Bit and LDPC scheme performs slightly better than Huffman and LDPC with an accuracy ranging from 35\% to 25\%. However, Student 1 and Student 2 still demonstrate better performance than the teacher, the baselines, and the conventional communications. Comparing Student 1 and Baseline 1, KD without model compression improves the accuracy by over 20\%, as the number of interfering users increases. Furthermore, Student 1 has a slightly better performance than Student 2, which shows that KD can reduce the performance loss while compressing the model.
\subsection{Ablation experiment}
\begin{table}[!ht]
  \renewcommand\arraystretch{0.8}
  \setlength{\tabcolsep}{5pt}
  \caption{Variations on BLEU score relative to the model trained with all losses.}
  \label{ablation}
  \centering
  \begin{tabular}{ccccc}
  \toprule
   &\multicolumn{2}{c}{Without interference}&\multicolumn{2}{c}{With interference}\\
   &SNR=6&SNR=15&SNR=6&SNR=15\\
   \midrule
   Without $\mathcal{L}_{distill}(\boldsymbol{\hat{s}}^\mathcal{T},\boldsymbol{\hat{s}}^\mathcal{S})$& -5.16\%& -2.25\%& -5.11\%& -1.07\%\\
   Without $\mathcal{L}_{distill}(\boldsymbol{x}^\mathcal{T},\boldsymbol{x}^\mathcal{S})$& \multirow{2}{*}{-6.08\%} & \multirow{2}{*}{-4.62\%}&\multirow{2}{*}{-5.73\%} & \multirow{2}{*}{-1.45\%}\\and $\mathcal{L}_{distill}(\boldsymbol{\hat{t}}^\mathcal{T},\boldsymbol{\hat{t}}^\mathcal{S})$& & \\
   Without $\mathcal{L}_{distill}(\boldsymbol{p}^\mathcal{T},\boldsymbol{p}^\mathcal{S})$ & \multirow{2}{*}{-5.08\%} & \multirow{2}{*}{-6.35\%} &\multirow{2}{*}{-0.89\%} & \multirow{2}{*}{-1.18\%}\\and $\mathcal{L}_{distill}(\boldsymbol{\hat{p}}^\mathcal{T},\boldsymbol{\hat{p}}^\mathcal{S})$& &\\
   Without $\mathcal{L}_{hard}(\boldsymbol{s}^\mathcal{S},\boldsymbol{\hat{s}}^\mathcal{S})$ & -18.63\%& -12.59\%&-11.71\% &-6.44\%\\
  \bottomrule
  \end{tabular}
\end{table}
In Table \ref{ablation}, we investigate the effect of various components of the loss function on the distilled student models. We categorize the distilled losses by considering the symmetric structure of the SemCom system. The SemCom systems with no interference and with one interference when SIR is 10 dB are considered. 

When there is no interference, the distilled knowledge from the outputs of the channel encoder and the semantic decoder has more impact on the performance than other distilled knowledge when there is strong noise in the system. Conversely, the knowledge from the intermediate outputs of the encoded semantic information $\boldsymbol{p}^\mathcal{T}$ and $\boldsymbol{p}^\mathcal{T}$ tends to have more influence on the performance when there is less noise. The cross-entropy loss $\mathcal{L}_{hard}(\boldsymbol{s}^\mathcal{S},\boldsymbol{\hat{s}}^\mathcal{S})$ is the most important loss function regardless of the presence of the interference because it has the hard label information, which is the direct way to improve the performance. When there is unseen interference, the effect of the hard labels information and the intermediate outputs of the semantic encoded information diminishes, while $\mathcal{L}_{distill}(\boldsymbol{\hat{s}}^\mathcal{T},\boldsymbol{\hat{s}}^\mathcal{S})$, $\mathcal{L}_{distill}(\boldsymbol{x}^\mathcal{T},\boldsymbol{x}^\mathcal{S})$ and $\mathcal{L}_{distill}(\boldsymbol{\hat{t}}^\mathcal{T},\boldsymbol{\hat{t}}^\mathcal{S})$ can have similar effects on the BLEU score to the condition with no interference. The student model improves the robustness of generalizing on the unseen data by learning from the teacher model through these distilled knowledge. Moreover, these results can be used as the guidance to determine the proportion of different loss functions.

\subsection{Complexity Analysis}
\begin{table}[!ht]
  \renewcommand\arraystretch{0.8}
  \caption{The complexity analysis with number of parameters, model size, training time and inference time.}
  \label{complexity}
  \centering
  \begin{tabular}{ccccc}
  \toprule 
  \multirow{2}{*}{}& \multirow{2}{*}{Parameters} & \multirow{2}{*}{\shortstack{Size\\ (MB)}} & \multirow{2}{*}{\shortstack{Training time\\ (ms/batch)}} & \multirow{2}{*}{\shortstack{Inference time\\(ms/sentence)}}\\ 
  & & & &\\
  \midrule
  Teacher & 2022672    & 12.46    & 108.86  & 24.10\\
  Student 1 & 2022672    & 12.46   & 199.75 & 22.90\\ 
  Student 2 & 1096976    & 6.98    & 169.87 & 14.87\\ 
  Student 3 & 946704     & 6.06    & 166.86 & 14.50\\ 
  Student 4 & 5376       & 0.05    & 166.86 & 14.18\\ 
  DeepSC    & 1462928    & 9.18    &  95.76 & 19.79\\
  LDPC & - & - & - & 42.74\\
  \bottomrule
  \end{tabular}
\end{table}
In Table \ref{complexity}, we conduct the complexity analysis for the student models in terms of the number of parameters, the size of the models, training time per batch and average inference time per sentence. With the model compression for semantic encoder-decoder components, the size of the semantic encoder and decoder is reduced from 12.46 MB to 6.98 MB. Moreover, about 50\% of the parameters are reduced for Student 2 compared with Student 1. Student 3 is compressed on the channel encoder and decoder, which has 0.92 MB, slimmer than Student 2. Furthermore, Student 4 uses post-training dynamic quantization based on Student 3, which further reduces the size from 6.06 MB to 0.05 MB. 

Furthermore, Student 1 costs about 199 ms/batch for training, which is the longest since it has the same size as the teacher but considers the extra distilled knowledge from the teacher. \hlbreakable{Students 2, 3, and 4 benefit from a reduced model size, resulting in approximately 30 ms/batch reduction in training time than Student 1. Despite this, they still require longer training time than the non-distilled models. This shows a potential drawback of KD for introducing extra computational overhead during model training.} In terms of the average inference time, it is related to the size of the model for the learning-based models. Students 2, 3, and 4 can have an inference time of less than 15 ms/sentence, whereas Teacher and Student 1 take more than 20 ms/sentence. This significantly outperforms the traditional communications system using LDPC. \hlbreakable{Therefore, KD-assisted model compression could reduce the sentence processing time during inference, improving the real-time latency. However, this comes at the expense of increasing training costs, which may not be suitable for applications, such as online learning, where the model is continuously trained with new incoming data.}

 \section{Conclusion} \label{sec5}
In this paper, we have proposed a KD-based SemCom system with MU interference. Specifically, four distilled student models have been designed and trained with the constraints of limited training samples. Performances have been compared and analyzed for different SNRs, SIRs, and the number of interference. Numerical results have shown distilled models perform better than the non-distilled baselines and the conventional communications system with Huffman codes and LDPC as the source and channel coding scheme when generalizing on unseen interference. KD can greatly improve the generalization and robustness of the student models. \hlbreakable{Moreover, the complexity analysis has been conducted to illustrate that KD can reduce inference time by compressing the model while compromising on training cost.} Furthermore, an ablation study has compared the importance of various distilled loss functions on the distilled student models. Finally, simulation results have also shown that the post-training dynamic quantization has a very limited effect on the system performance. 


\ifCLASSOPTIONcaptionsoff
  \newpage
\fi



\begin{IEEEbiographynophoto}{Chenguang Liu}
Chenguang Liu received his B.E. degree in software engineering from Dalian University of Technology, Dalian, P.R.China, in 2016 and M.S. degree in advanced computer science from The University of Manchester, U.K., in 2017. He is currently studying as a Ph.D. student at the University of Warwick, U.K. His research interests include deep learning, wireless communications and cooperative perception.
\end{IEEEbiographynophoto}
\begin{IEEEbiographynophoto}{Yuxin Zhou}
Yuxin Zhou received his B.E. degree in robotics engineering from Harbin Institute of Technology, Weihai, P.R.China, in 2020. He is currently studying as a M.E. student at the Southern University of Technology and Science, Shenzhen, P.R.China. His research interests include deep learning and computer vision.
\end{IEEEbiographynophoto}
\begin{IEEEbiographynophoto}{Yunfei Chen}
Yunfei Chen (S'02-M'06-SM'10) received his B.E.and M.E. degrees in electronics engineering from Shanghai Jiaotong University, Shanghai, P.R.China,in 1998 and 2001, respectively. He received his Ph.D. degree from the University of Alberta in 2006. He is currently working as a Professor at the University of Durham, U.K. His research interests include wireless communications, cognitive radios, wireless relaying and energy harvesting.
\end{IEEEbiographynophoto}
\begin{IEEEbiographynophoto}{Shuang-Hua Yang}
Shuang-Hua Yang (Senior Member, IEEE) received the B.S. degree in instrument and automation and the M.S. degree in process control from the China University of Petroleum, Beijing, China, in 1983 and 1986, respectively, and the Ph.D. degree in intelligent systems from Zhejiang University, Hangzhou, China, in 1991. He is currently the director of the Shenzhen Key Laboratory of Safety and Security for Next Generation of Industrial Internet, Southern University of Science and Technology, and a Professor and the Head of the Department of Computer Science, University of Reading, U.K. His research interests include Internet of Things, industrial internet, Cyber-Physical System safety and security, Process Systems Engineering. Prof. Yang is a fellow of IET and InstMC, U.K. He is an Associate Editor of IET Cyber-Physical Systems: Theory and Applications.
\end{IEEEbiographynophoto}

\end{document}